\makeatletter

\font\tenmsx=msxm10
\font\sevenmsx=msxm7
\font\fivemsx=msxm5
\font\tenmsy=msym10
\font\sevenmsy=msym7
\font\fivemsy=msym5
\newfam\msxfam
\newfam\msyfam
\textfont\msxfam=\tenmsx  \scriptfont\msxfam=\sevenmsx
  \scriptscriptfont\msxfam=\fivemsx
\textfont\msyfam=\tenmsy  \scriptfont\msyfam=\sevenmsy
  \scriptscriptfont\msyfam=\fivemsy

\def\hexnumber@#1{\ifnum#1<10 \number#1\else
 \ifnum#1=10 A\else\ifnum#1=11 B\else\ifnum#1=12 C\else
 \ifnum#1=13 D\else\ifnum#1=14 E\else\ifnum#1=15 F\fi\fi\fi\fi\fi\fi\fi}

\def\msx@{\hexnumber@\msxfam}
\def\msy@{\hexnumber@\msyfam}
\mathchardef\boxdot="2\msx@00
\mathchardef\boxplus="2\msx@01
\mathchardef\boxtimes="2\msx@02
\mathchardef\square="0\msx@03
\mathchardef\blacksquare="0\msx@04
\mathchardef\centerdot="2\msx@05
\mathchardef\lozenge="0\msx@06
\mathchardef\blacklozenge="0\msx@07
\mathchardef\circlearrowright="3\msx@08
\mathchardef\circlearrowleft="3\msx@09
\mathchardef\rightleftharpoons="3\msx@0A
\mathchardef\leftrightharpoons="3\msx@0B
\mathchardef\boxminus="2\msx@0C
\mathchardef\Vdash="3\msx@0D
\mathchardef\Vvdash="3\msx@0E
\mathchardef\vDash="3\msx@0F
\mathchardef\twoheadrightarrow="3\msx@10
\mathchardef\twoheadleftarrow="3\msx@11
\mathchardef\leftleftarrows="3\msx@12
\mathchardef\rightrightarrows="3\msx@13
\mathchardef\upuparrows="3\msx@14
\mathchardef\downdownarrows="3\msx@15
\mathchardef\upharpoonright="3\msx@16

\mathchardef\downharpoonright="3\msx@17
\mathchardef\upharpoonleft="3\msx@18
\mathchardef\downharpoonleft="3\msx@19
\mathchardef\rightarrowtail="3\msx@1A
\mathchardef\leftarrowtail="3\msx@1B
\mathchardef\leftrightarrows="3\msx@1C
\mathchardef\rightleftarrows="3\msx@1D
\mathchardef\Lsh="3\msx@1E
\mathchardef\Rsh="3\msx@1F
\mathchardef\rightsquigarrow="3\msx@20
\mathchardef\leftrightsquigarrow="3\msx@21
\mathchardef\looparrowleft="3\msx@22
\mathchardef\looparrowright="3\msx@23
\mathchardef\circeq="3\msx@24
\mathchardef\succsim="3\msx@25
\mathchardef\gtrsim="3\msx@26
\mathchardef\gtrapprox="3\msx@27
\mathchardef\multimap="3\msx@28
\mathchardef\therefore="3\msx@29
\mathchardef\because="3\msx@2A
\mathchardef\doteqdot="3\msx@2B

\mathchardef\triangleq="3\msx@2C
\mathchardef\precsim="3\msx@2D
\mathchardef\lesssim="3\msx@2E
\mathchardef\lessapprox="3\msx@2F
\mathchardef\eqslantless="3\msx@30
\mathchardef\eqslantgtr="3\msx@31
\mathchardef\curlyeqprec="3\msx@32
\mathchardef\curlyeqsucc="3\msx@33
\mathchardef\preccurlyeq="3\msx@34
\mathchardef\leqq="3\msx@35
\mathchardef\leqslant="3\msx@36
\mathchardef\lessgtr="3\msx@37
\mathchardef\backprime="0\msx@38
\mathchardef\risingdotseq="3\msx@3A
\mathchardef\fallingdotseq="3\msx@3B
\mathchardef\succcurlyeq="3\msx@3C
\mathchardef\geqq="3\msx@3D
\mathchardef\geqslant="3\msx@3E
\mathchardef\gtrless="3\msx@3F
\mathchardef\sqsubset="3\msx@40
\mathchardef\sqsupset="3\msx@41
\mathchardef\trianglerighteq="3\msx@44
\mathchardef\trianglelefteq="3\msx@45
\mathchardef\bigstar="0\msx@46
\mathchardef\between="3\msx@47
\mathchardef\blacktriangledown="0\msx@48
\mathchardef\blacktriangleright="3\msx@49
\mathchardef\blacktriangleleft="3\msx@4A
\mathchardef\blacktriangle="0\msx@4E
\mathchardef\triangledown="0\msx@4F
\mathchardef\eqcirc="3\msx@50
\mathchardef\lesseqgtr="3\msx@51
\mathchardef\gtreqless="3\msx@52
\mathchardef\lesseqqgtr="3\msx@53
\mathchardef\gtreqqless="3\msx@54
\mathchardef\Rrightarrow="3\msx@56
\mathchardef\Lleftarrow="3\msx@57
\mathchardef\veebar="2\msx@59
\mathchardef\barwedge="2\msx@5A
\mathchardef\doublebarwedge="2\msx@5B
\mathchardef\angle="0\msx@5C
\mathchardef\measuredangle="0\msx@5D
\mathchardef\sphericalangle="0\msx@5E
\mathchardef\varpropto="3\msx@5F
\mathchardef\smallsmile="3\msx@60
\mathchardef\smallfrown="3\msx@61
\mathchardef\Subset="3\msx@62
\mathchardef\Supset="3\msx@63
\mathchardef\Cup="2\msx@64

\mathchardef\Cap="2\msx@65

\mathchardef\curlywedge="2\msx@66
\mathchardef\curlyvee="2\msx@67
\mathchardef\leftthreetimes="2\msx@68
\mathchardef\rightthreetimes="2\msx@69
\mathchardef\subseteqq="3\msx@6A
\mathchardef\supseteqq="3\msx@6B
\mathchardef\bumpeq="3\msx@6C
\mathchardef\Bumpeq="3\msx@6D
\mathchardef\lll="3\msx@6E

\mathchardef\ggg="3\msx@6F

\mathchardef\circledS="0\msx@73
\mathchardef\pitchfork="3\msx@74
\mathchardef\dotplus="2\msx@75
\mathchardef\backsim="3\msx@76
\mathchardef\backsimeq="3\msx@77
\mathchardef\complement="0\msx@7B
\mathchardef\intercal="2\msx@7C
\mathchardef\circledcirc="2\msx@7D
\mathchardef\circledast="2\msx@7E
\mathchardef\circleddash="2\msx@7F
\def\ulcorner{\delimiter"4\msx@70\msx@70 }
\def\urcorner{\delimiter"5\msx@71\msx@71 }
\def\llcorner{\delimiter"4\msx@78\msx@78 }
\def\lrcorner{\delimiter"5\msx@79\msx@79 }
\def\yen{\mathhexbox\msx@55 }
\def\checkmark{\mathhexbox\msx@58 }
\def\circledR{\mathhexbox\msx@72 }
\def\maltese{\mathhexbox\msx@7A }
\mathchardef\lvertneqq="3\msy@00
\mathchardef\gvertneqq="3\msy@01
\mathchardef\nleq="3\msy@02
\mathchardef\ngeq="3\msy@03
\mathchardef\nless="3\msy@04
\mathchardef\ngtr="3\msy@05
\mathchardef\nprec="3\msy@06
\mathchardef\nsucc="3\msy@07
\mathchardef\lneqq="3\msy@08
\mathchardef\gneqq="3\msy@09
\mathchardef\nleqslant="3\msy@0A
\mathchardef\ngeqslant="3\msy@0B
\mathchardef\lneq="3\msy@0C
\mathchardef\gneq="3\msy@0D
\mathchardef\npreceq="3\msy@0E
\mathchardef\nsucceq="3\msy@0F
\mathchardef\precnsim="3\msy@10
\mathchardef\succnsim="3\msy@11
\mathchardef\lnsim="3\msy@12
\mathchardef\gnsim="3\msy@13
\mathchardef\nleqq="3\msy@14
\mathchardef\ngeqq="3\msy@15
\mathchardef\precneqq="3\msy@16
\mathchardef\succneqq="3\msy@17
\mathchardef\precnapprox="3\msy@18
\mathchardef\succnapprox="3\msy@19
\mathchardef\lnapprox="3\msy@1A
\mathchardef\gnapprox="3\msy@1B
\mathchardef\nsim="3\msy@1C
\mathchardef\napprox="3\msy@1D
\mathchardef\nsubseteqq="3\msy@22
\mathchardef\nsupseteqq="3\msy@23
\mathchardef\subsetneqq="3\msy@24
\mathchardef\supsetneqq="3\msy@25
\mathchardef\subsetneq="3\msy@28
\mathchardef\supsetneq="3\msy@29
\mathchardef\nsubseteq="3\msy@2A
\mathchardef\nsupseteq="3\msy@2B
\mathchardef\nparallel="3\msy@2C
\mathchardef\nmid="3\msy@2D
\mathchardef\nshortmid="3\msy@2E
\mathchardef\nshortparallel="3\msy@2F
\mathchardef\nvdash="3\msy@30
\mathchardef\nVdash="3\msy@31
\mathchardef\nvDash="3\msy@32
\mathchardef\nVDash="3\msy@33
\mathchardef\ntrianglerighteq="3\msy@34
\mathchardef\ntrianglelefteq="3\msy@35
\mathchardef\ntriangleleft="3\msy@36
\mathchardef\ntriangleright="3\msy@37
\mathchardef\nleftarrow="3\msy@38
\mathchardef\nrightarrow="3\msy@39
\mathchardef\nLeftarrow="3\msy@3A
\mathchardef\nRightarrow="3\msy@3B
\mathchardef\nLeftrightarrow="3\msy@3C
\mathchardef\nleftrightarrow="3\msy@3D
\mathchardef\divideontimes="2\msy@3E
\mathchardef\varnothing="0\msy@3F
\mathchardef\nexists="0\msy@40
\mathchardef\mho="0\msy@66
\mathchardef\thorn="0\msy@67
\mathchardef\beth="0\msy@69
\mathchardef\gimel="0\msy@6A
\mathchardef\daleth="0\msy@6B
\mathchardef\lessdot="3\msy@6C
\mathchardef\gtrdot="3\msy@6D
\mathchardef\ltimes="2\msy@6E
\mathchardef\rtimes="2\msy@6F
\mathchardef\shortmid="3\msy@70
\mathchardef\shortparallel="3\msy@71
\mathchardef\smallsetminus="2\msy@72
\mathchardef\thicksim="3\msy@73
\mathchardef\thickapprox="3\msy@74
\mathchardef\approxeq="3\msy@75
\mathchardef\succapprox="3\msy@76
\mathchardef\precapprox="3\msy@77
\mathchardef\curvearrowleft="3\msy@78
\mathchardef\curvearrowright="3\msy@79
\mathchardef\digamma="0\msy@7A
\mathchardef\varkappa="0\msy@7B
\mathchardef\hslash="0\msy@7D
\mathchardef\hbar="0\msy@7E
\mathchardef\backepsilon="3\msy@7F
\def\Bbb{\ifmmode\let\next\Bbb@\else
 \def\next{\errmessage{Use \string\Bbb\space only in math mode}}\fi\next}
\def\Bbb@#1{{\Bbb@@{#1}}}
\def\Bbb@@#1{\fam\msyfam#1}

\def\inv{^{\raise.15ex\hbox{${
  \scriptscriptstyle -}$}\kern-.05em 1}}

\def\Dsl{\,\raise.15ex\hbox{$/$}\mkern-13.5mu D}
\def\dsl{\raise.15ex\hbox{$/$}\kern-.57em\hbox{$\partial$}}

\def\lspace{\ifx\answ\bigans{}\else\qquad\fi}

\def\CR{\hbox{{$\cal R$}}}

\def\lform{\hbox{$\sqcup$}\llap{\hbox{$\sqcap$}}}
\def\darr#1{\raise1.5ex\hbox{$\leftrightarrow$}
\mkern-16.5mu #1}

\def\INT{{\textstyle \int\kern-.642em\int}}

\def\eps{{\epsilon}}

\def\small{\scriptstyle}
\def\tens{\mathop{\otimes}}

\def\id{{\rm id}}

\def\eqn#1#2{\begin{equation}#2\label{#1}\end{equation}}

\def\o{{}_{(1)}}\def\t{{}_{(2)}}

\def\text#1{\mbox{\rm #1}}
\def\note#1{}

\def\blacksquare{{\lform}}
\def\frac#1#2{{{#1\over#2}}}

\def\goth#1{{#1}}

\def\align#1{\begin{eqnarray*}#1\end{eqnarray*}}

\def\vect{{\bf t}}\def\vecu{{\bf
u}}

\def\difid{{\goth M}}
\def\vect{{\bf t}}
\def\vecu{{\bf u}}

\def\tr{{\rm tr}}

\makeatother

\documentstyle{article}
\begin{document}
\newtheorem{prop}{Proposition}[section]
\newtheorem{lemma}[prop]{Lemma}
\newtheorem{thm}[prop]{Theorem}
\newtheorem{df}[prop]{Definition}
\newtheorem{rk}[prop]{Remark}
\newtheorem{cor}[prop]{Corollary}
\newtheorem{ex}[prop]{Example}
\baselineskip 12pt
\title{\marginpar{\vskip -.5in \hskip -1in \small DAMTP/92-33}
A class of bicovariant differential calculi on Hopf algebras}

\author{Tomasz Brzezi\'{n}ski \thanks{Supported by St. John's College,
Cambridge \& KBN grant 2 0218 91 01} \& Shahn Majid \thanks{SERC
Fellow and Drapers Fellow of Pembroke College, Cambridge}\\
Department of Applied Mathematics \& Theoretical Physics \\
University of Cambridge \\ CB3 9EW, U.K.}
\date{July 1992}

\maketitle

\begin{quote}ABSTRACT We introduce a large class of bicovariant
differential calculi on any quantum group $A$, associated to
$Ad$-invariant elements. For example, the deformed trace element on
$SL_q(2)$ recovers Woronowicz' $4D_\pm$ calculus. More generally, we
obtain a sequence of differential calculi on each quantum group
$A(R)$, based on the theory of the corresponding braided groups $B(R)$.
Here $R$ is any regular solution of the QYBE.
\end{quote}

\baselineskip 20pt

\section{Introduction}

The differential geometry of quantum groups  was introduced by S.L.
Woronowicz in \cite{woronowicz1} for $SU_q(2)$ and then formulated for
any compact matrix quantum group in \cite{woronowicz3}.  This theory is
based on the idea that the differential and (co)algebraic structures of
$A$ should be compatible in the sense that $A$ can coact covariantly
on the algebra of its differential forms. This idea leads to the
notions of right-, left- and bi-covariant differential calculi over
$A$. The last of these fully respects the (co)algebraic structure of
$A$ and therefore seems to be the most natural one. By now, the
importance of this notion in physics has become clear and several
examples are known, constructed by hand. In the present paper our goal
is to show how to obtain such calculi more systematically.

First, let us note that the bicovariance condition, however
restrictive, does not determine the differential structure of $A$
uniquely so that there is no preferred such structure. Some attempts
have been undertaken \cite{manin1} \cite{maltsiniotis1} \cite{sudbery1}
\cite{sudbery5} in order to decrease the number of allowed differential
calculi over $A$ by means of some additional requirements, e.g. that
the differential structure of $A$ should be obtained from the
differential structure on a quantum space on which $A$ acts
covariantly. Such considerations can help us characterise uniquely and
hence construct the desired differential calculus.  So far,
this kind of approach has proven successful in obtaining examples for
$GL_q(N)$ and some other matrix quantum groups.

There is another method of construction of bicovariant differential
calculi due to Jurco\cite{jurco1}, which works for the standard quantum
groups associated to simple Lie algebras.  It is based on the
observation that the space of left-invariant vector fields should
correspond to the quantum universal enveloping algebra $U_q(g)$ dual to
$A$. The correspondence is not quite direct. As noted in
\cite{woronowicz3} left invariant vector fields are induced by certain
linear functionals on $A$ vanishing on an ideal (which we will denote by
$\difid$)
defined by the differential calculus.
{}From these are obtained related linear functionals that can then be
used to determine
commutation relations  between 1-forms and elements of $A$.  The latter
functionals can be built up from the functionals $l_\pm$\cite{frt1} in
the FRT description of $U_q(g)$. Therefore, reversing these arguments
one can use $l_\pm$ to construct vector fields and to obtain the
relevant differential calculus (cf.  \cite{zumino1}). This method gives
a special class of bicovariant differential calculi on the standard
quantum groups. See also the preprint \cite{SWZ} for $SL_q(N)$, received
after the present work was completed.

By contrast to such specific methods, we introduce in the present paper
a general construction for bicovariant differential calculi based in
the abstract theory of Hopf algebras. In our construction we associate
a bicovariant differential calculus to every element $\alpha$ of the
Hopf algebra that is invariant under the adjoint coaction. For example,
the calculi described in \cite{jurco1} are related to the invariant
quantum trace of a quantum matrix. It turns out that all
previously-known examples of bicovaraint differential calculi can be
similarly obtained by our general construction. On the other hand, our
construction is not limited either to matrix quantum groups or to the
standard quantum groups associated to simple Lie algebras. Moreover,
because the construction is not tied to specific applications, we
obtain more novel and unexpected calculi even for the standard quantum
groups. This is somewhat analogous to the commutative case, where the
axioms of a differential structure admit exotic solutions in addition to
the standard differential structures, as known even for $\Bbb R^4$. Several
examples are collected at the end of Section~3 after proving our main
theorem. A novel feature of our approach is that the differential calculi
in our construction form a kind of algebra given by addition and multiplication
of the underlying $Ad$-invariant elements.

We use the following notation: $ A$ is a Hopf algebra over the
field $k$ (such as $\Bbb R$ or $\Bbb C$), with coproduct $\Delta: {A}
\rightarrow {A} \otimes A$, counit $\epsilon: A
\rightarrow k$ and antipode $S:  A \rightarrow A$.

\section{Differential structures on quantum groups}

 We recall first some basic notions about differential calculus on
quantum groups, according to the general theory formulated in
\cite{woronowicz3} (cf \cite{brzezinski3}, \cite{zumino1}). The section
then introduces more technical facts needed for our construction.  To
begin with, one says that $(\Gamma ,d)$ is a {\it first order
differential calculus} over a Hopf algebra ($A,\Delta,S,\epsilon$) if
$d: A \rightarrow \Gamma$ is a linear map obeying the Leibniz rule,
$\Gamma$ is a bimodule over $A$ and every element of $\Gamma$ is of the
form $\sum_{k} a_{k}db_{k}$, where $a_{k}, b_{k} \in A$.  It is known
that every differential calculus on an algebra $ A$ can be obtained as
a quotient of a universal one $(A^{2}, D)$, where $A ^{2} = {\rm ker}
\mu$ ($\mu : A
 \otimes  A \rightarrow A$ is the multiplication map in $A$) and  $D: A
\rightarrow A ^{2}$ is defined by \begin{equation} Da = a \otimes 1 - 1
\otimes a.  \end{equation}

The map $D$ obeys Leibniz rule provided $A ^{2}$ has an $A$
bimodule structure given by
\begin{eqnarray}
c(\sum_{k} a_{k} \otimes b_{k}) = \sum_{k} ca_{k} \otimes b_{k},\qquad
(\sum_{k} a_{k} \otimes b_{k} )c = \sum_{k} a_{k} \otimes b_{k}c
\label{eq:module}
\end{eqnarray}
for any $\sum_{k} a_{k} \otimes b_{k} \in {A}^{2}$, $c \in A$.
Furthermore every element of $A^{2}$ can be represented in the form
$\sum_{k} a_{k} Db_{k}$. Indeed, let $\rho = \sum_{k} a_{k} \otimes b_{k} \in
A^{2}$. Then
$$\rho = - \sum_{k} a_{k} (b_{k} \otimes 1 - 1 \otimes b_{k}) + \sum_{k}
a_{k}b_{k} \otimes 1 = -\sum_{k} a_{k} D b_{k}$$

Hence $(A^{2} , D)$ is a first order differential calculus
over $A$ as stated. Any first order differential calculus on
$A$ can be realized as $(\Gamma = A^{2} / N$, $d = \pi D)$
where $N \subset A^{2}$ is a sub-bimodule of $A^{2}$ and $\pi : A^{2}
\rightarrow A^{2} /N$ is a
canonical epimorphism.

In the analysis of differential structures over quantum groups (Hopf algebras)
a crucial r\^ole is played by the covariant differential calculi.

\begin{df} Let $(\Gamma , d)$ be a first order differential calculus
over a Hopf
algebra $A$. We say that $(\Gamma , d )$ is:
\begin{enumerate}
\item {\em left-covariant} if for any $a_k,b_k\in A$,
$$ (\sum a_{k}db_{k} = 0) \Rightarrow (\sum \Delta(a_{k})(id \otimes d)
\Delta(b_{k}) = 0).$$
\item {\em right-covariant} if for any $a_k,b_k\in A$,
$$ (\sum a_{k}db_{k} = 0) \Rightarrow (\sum \Delta(a_{k})(d \otimes id)
\Delta(b_{k}) = 0).$$
\item {\em bicovariant} if it is left- and right-covariant.
\end{enumerate}
\end{df}

The notion of a covariant differential calculus leads to:
\begin{df}
Let $A$ be a Hopf algebra and $\Gamma$ be an $A$-bimodule as well as
a left $A$-comodule (right $A$-comodule) with
coaction $\Delta _{L}$ ($\Delta _{R}$ resp.) Then we say that $\Gamma$
is a {\em left-covariant (right-covariant) $A$-bimodule of} if:
$${\Delta _{L}}_{(R)} (a\rho) = (\Delta a) \Delta _{L(R)} \rho ,
\hspace{1cm} \Delta _{L(R)} (\rho a) = (\Delta _{L(R)} \rho ) \Delta
a$$
for any $a \in A$ and $\rho \in \Gamma$. It is {\em bicovariant} if both
conditions hold.

Let $\Gamma_{1}, \; \Gamma_{2}$ be two bicovariant $A$-bimodules with
coactions  $\Delta _{1L(R)}$ and $\Delta _{2L(R)}$
respectively. We say that the linear map $\phi: \Gamma_{1} \rightarrow
\Gamma _{2}$ is a {\em bicovariant bimodule map} if:
$$(id \otimes \phi )\Delta _{1L} = \Delta _{2L} \phi, \hspace{1cm}
(\phi \otimes id) \Delta _{1R} = \Delta _{2R} \phi$$
\end{df}

Left-covariance (resp. right-covariance) of a first order differential
calculus $(\Gamma ,d)$ implies that $\Gamma$ is a
left-covariant (resp. right-covariant) $A$-bimodule and that $d$ is a
bicovariant bimodule map, i.e. if $\Delta _{L} : \Gamma
\rightarrow A \otimes \Gamma $, (resp. $\Delta_{R}: \Gamma \rightarrow
\Gamma \otimes A$) is a left coaction (resp. right coaction) then
\begin{equation}
\Delta_{L}d = (id \otimes d) \Delta ,\quad (\Delta _{R} d = (d \otimes
id) \Delta)
\label{four}
\end{equation}

If $(\Gamma , d)$ is a bicovariant differential calculus over $A$ then
$\Gamma$ is a bicovariant
$A$-bimodule.  Bicovariancy of $(\Gamma ,d)$ allows one to
define a wedge product of forms, construct  the exterior (${\bf
Z}_{2}$-graded) algebra
$\Omega({A})$ (with $\Omega^{1}({A}) =
\Gamma$) and
extend the differential $d$ uniquely to the whole of $\Omega(A)$.
Here $d$ obeys  the graded Leibniz rule, $d^{2}
= 0$ and equations (\ref{four}) are obeyed also for the extensions of
coactions to the whole of $\Omega({A})$.

 The exterior product in $\Omega(A)$ is defined by the
following construction. We say that an element $\omega \in
\Omega^{1}(A)$ is
{\em left-invariant} ({\em right-invariant}) if $\Delta _{L}(\omega) = 1
\otimes \omega$ ($\Delta _{R}(\omega) = \omega \otimes 1$). A one-form
$\omega$ is {\em bi-invariant} if it is left- and right-invariant.
Given a basis $\{ \omega_i \}$ of the space of all left-invariant
 1-forms any
element $\rho \in \Omega^{1}(A)$ can be represented uniquely
as $\rho = \sum a_{i} \omega _{i}$, where $a_{i} \in A$. Similarly any
element $\rho$ of $\Omega^{1}(A) \otimes \Omega^{1}(A)$ can be
represented as $\rho =\sum a_{ij}
\omega _{i} \otimes \eta _{j}$, where $a_{ij} \in A$, $\omega
_{i}$ are left-invariant and $\eta _{j}$ are right-invariant. Now
define $\Omega^{2}(A) =
\Omega^{1}(A) \otimes \Omega^{1}(A) / {\rm ker}(id-\sigma)$,
where $\sigma :\Omega^{1}(A) \otimes \Omega^{1}(A)
\rightarrow \Omega^{1}(A) \otimes \Omega^{1}(A)$, is such that
$\sigma: \omega_{1} \otimes \omega_{2} \mapsto
 \omega_{2} \otimes \omega_{1}$ for any
left-invariant $\omega_{1}$ and right-invariant $\omega_{2}$.
This definition extends to $\Omega ^{n}(A)$ for any positive n.
The exterior differentiation $d$ is defined with the help of a
bi-invariant one-form $\theta$ in such a way that
\begin{equation}
d \rho = \theta \wedge \rho - (-1)^{\partial \rho} \rho \wedge \theta
\label{six}
\end{equation}
 for any homogeneous $\rho
\in \Omega(A)$ of degree $\partial \rho$.

Bicovariance of the differential calculus $(\Omega^{1}(A), d)$
means not only that $\Omega ^{1}(A)$ is a bicovariant $
A$-bimodule and that higher modules $\Omega^{n}(A)$ can be
defined but also that $\Omega(A)$ is an exterior
bialgebra (see \cite{brzezinski3}), i.e. $\Omega(A)$ is a $\Bbb Z_2$-graded
or super bialgebra and the graded-coproduct $\Delta $ in $\Omega(A)$ is
compatibile with $d$. The tensor product $\Omega(A)
\otimes \Omega(A)$ is graded in a natural way according to
$$\Omega(A) \otimes \Omega(A) =
\bigoplus_{n=0}^{\infty} \bigoplus_{k=0}^{n}\Omega^{k}(A) \otimes
\Omega^{n-k}(A).$$
The action of $d$ on this tensor product is defined by means of the
graded Leibniz rule. The graded-coproduct $\Delta$ on $\Omega^1(A)$
is given by
\begin{equation}
\Delta = \Delta_R + \Delta _L
\label{one}
\end{equation}
and then extended to the whole of exterior algebra by
\[ \Delta (a_0da_1 \cdots da_n) = \Delta (a_0) \Delta (da_1) \cdots
\Delta (da_n). \]

The universal calculus $(A^{2}, D)$ is probably the most
important example of a bicovariant differential calculus over a general
Hopf algebra $A$. The exterior  algebra related to it will be
denoted by $\Lambda A$. This $\Lambda A$ is a quotient of the unital
differential envelope of $A$, $\Omega A$ (cf \cite{coq-kast1},
\cite{kastler1}) by the kernel of the symmetrizer $id - \sigma$. The
comodule actions $\Delta_{L}^{U}, \;
\Delta_{R}^{U}$ of $A$ on
$A^2$ are given by
$$\Delta_{L}^{U}(\sum a_{k} \otimes b_{k}) = \sum a_{k (1)}b_{k(1)} \otimes
a_{k(2)} \otimes b_{k(2)}$$
$$\Delta_{R}^{U}(\sum a_{k} \otimes b_{k}) = \sum a_{k(1)} \otimes
b_{k(1)} \otimes a_{k(2)}b_{k (2)}$$
Here we have used standard notation\cite{sweedler1}
$\Delta a = \sum a_{(1)} \otimes a_{(2)}$. Furthermore $\Omega A$ is
an exterior bialgebra with graded-coproduct (\ref{one}).

Let $(\Gamma ,d)$ be a bicovariant differential calculus over a Hopf
algebra $A$ and let $N \subset A^2$ be such that $\Gamma = A^2 / N$.
The maps $\Delta_L^U$, $\Delta_R^U$ defined above allow one to
describe $N$ in terms of a right ideal $\difid \subset \ker \epsilon$.
The latter can be defined by the relation
\[A \otimes \difid  = (id \otimes \epsilon \otimes id)\Delta_L^U N. \]
This is equivalent to
\[\difid  \otimes A = (\epsilon \otimes id \otimes \id)\Delta_R^UN \]
because the differential calculus is bicovariant. The ideal $\difid $ has
an additional property, namely it is an $Ad_R$-invariant subspace of $A$. To
be more precise
let us recall the notion of the right adjoint coaction of a
Hopf algebra $A$ on itself. This is the
linear map $Ad_{R}: A \rightarrow A \otimes A$ given by
$$Ad_{R} (a) = \sum a_{(2)} \otimes (Sa_{(1)})a_{(3)},$$
for any $a \in A$ (here $(id \otimes \Delta)\Delta a = \sum
a_{(1)} \otimes a_{(2)} \otimes a_{(3)}$).  We say that an element
$\alpha \in A$ is
{\em $Ad_{R}$-invariant} if:
\[ Ad_R (\alpha) = \alpha \otimes 1. \]
Similarly we say that a subspace $B \subset A$ is {\em
$Ad_R$-invariant} if $Ad_{R}(B)\subset B \otimes A$. In our case the ideal
$\difid $ is such that
\[ Ad_R (\difid ) \subset \difid  \otimes A . \]

We note that the map $(id \otimes \epsilon \otimes id) \Delta_R^U$ is
an isomorphism of $A$-bimodules. This implies that $\difid $ defines
$N$ uniquely, hence $\difid $ and $N$ can be treated equivalently with
respect to the calculus they define.

In the case of the universal calculus we define a map $\omega_U : A
\rightarrow A^2$ by
\begin{equation}
\omega_U(a) = \sum Sa_{(1)}Da_{(2)} = \epsilon (a) 1 \otimes 1 - \sum
Sa_{(1)} \otimes a_{(2)}.
\end{equation}
This map, being the quantum group equivalent of the Maurer-Cartan form,
will play an important role in the construction of bicovariant
differential calculi on $A$. It is easy to check that $\omega_U(a)$ is
a left-invariant 1-form for any $a \in A$ and that $\omega_U$
intertwines $\Delta^U_R$ and $Ad_R$.

We begin by showing how to use $\omega_U$ to
reconstruct the universal differential $D$ on $A$. Let us recall (cf.
\cite{sweedler1}) that an
element $\Lambda \in A$ is called a right-integral in $A$ if $\Lambda
a = \epsilon(a) \Lambda$ for any $a \in A$. We have the following:
\begin{prop}
Let $A$ be a Hopf algebra and $\Lambda \in A$ a right
integral in $A$ such that $\epsilon (\Lambda ) \neq 0$. Put
$\tilde {\theta} = -\frac{1}{\epsilon (\Lambda)}
\omega_{U} (\Lambda)$. Then
\begin{equation}
Da = \tilde {\theta} a - a\tilde {\theta}
\label{twelve}
\end{equation}
for any $a \in A$.
\end{prop}
{\bf Proof.} We compute the left hand side of (\ref{twelve})
directly,
\begin{eqnarray*}
\tilde {\theta} a - a \tilde {\theta} &=&- \frac{1}{\epsilon(\Lambda )}
(\epsilon (\Lambda
) 1\otimes a - a \otimes 1 \epsilon (\Lambda ) + (S\Lambda _{(1)})
\otimes \Lambda _{(2)} a - a (S\Lambda _{(1)}) \otimes \Lambda _{(2)})
\\
&=& Da + \frac{1}{\epsilon (\Lambda )}(a (S\Lambda _{(1)}) \otimes
\Lambda _{(2)} - a_{(0)} (Sa_{(1)}) (S\Lambda _{(1)}) \otimes \Lambda
_{(2)} a_{(2)}) \\
&=& Da + \epsilon (\Lambda )^{-1} (a (S \Lambda _{(1)}) \otimes
\Lambda _{(2)} - a_{(0)} S(\Lambda _{(1)} a_{(1)}) \otimes (\Lambda
a)_{(2)}) \\
&=& Da + \epsilon (\Lambda )^{-1} (a (S\Lambda _{(1)}) \otimes \Lambda
_{(2)} - a (S\Lambda _{(1)}) \otimes \Lambda _{(2)}) \\
&=& Da
\end{eqnarray*}
as required. \hspace{1cm} $\Box$

\vspace{20pt}

We notice here that the 1-form $\tilde {\theta}$ of Proposition~2.3 is
not necessarily bi-invariant. This implies further that $\tilde
{\theta}$ cannot necessarily be extended to a universal exterior
derivative $d$ via equation (5) above. However, we see that a
sufficient condition for $\tilde\theta$ to be bi-invariant is that
$\Lambda$ is $Ad_R$-invariant and that in this case we will obtain a
bicovariant calculus and an exterior algebra. This observation is a
leading idea behind our general construction in the next section.

\section{Construction  of bicovariant differential calculi on quantum
groups} As explained in the previous section, bicovariancy of the first
order differential calculus $(\Omega^{1}({A}),d)$ implies that
$\Omega^{1}({ A})$ is a bicovariant bimodule over $ A$ and that higher
modules $\Omega ^{n} (A)$ can be defined. The problem remains how to
construct examples of such $(\Omega^1(A) ,d)$. We do this now by
extending our reconstruction of the universal calculus in Proposition~
2.3 to the case of a calculus induced by an arbitrary $Ad_R$-invariant
element.

The main result of the section is:

\begin{thm}
Let $A$ be a Hopf algebra and let $\alpha \in A$ ($\alpha \neq 0,1$)
be an $Ad_{R}$-invariant element of $A$. Then there exists bicovariant
differential calculus $(\Gamma _{\alpha} ,d)$ such that
\begin{equation}
da = \lambda ^{-1} ((\pi_\alpha \circ\omega _{U} (\alpha))a -
a(\pi_\alpha \circ\omega _{U}
(\alpha)))
\label{cov.def.1}
\end{equation}
for any $a \in A$, where $\pi_\alpha : A^{2} \rightarrow \Gamma
_{\alpha}$ is a natural projection and $\lambda \in k^{*}$.
\label{cov.prop}
\end{thm}
{\bf Proof.} We consider the sub-bimodule $N_{\alpha} = \{ bn_{\alpha}(a)c
: a,b,c \in A \}$ of $A^{2}$ where
$$n_{\alpha} (a) = (\epsilon (\alpha) + \lambda ) (a \otimes 1 - 1
\otimes a) + \sum S\alpha _{(1)} \otimes \alpha _{(2)} a - \sum a
S\alpha _{(1)} \otimes \alpha _{(2)},$$
 and define
$$\Gamma _{\alpha} \equiv A^{2} / N_{\alpha}.$$
We have to prove that $\Gamma _{\alpha}$ is a bicovariant bimodule of
$A^{2}$ and that $d = \pi_\alpha \circ D$ is given by the equation
(\ref{cov.def.1}). To prove the former statement we need to show that
\begin{equation}
\Delta _{L} ^{U} N _{\alpha} \subset A \otimes N _{\alpha} ,
\hspace{1cm} \Delta _{R}^{U} N_{\alpha} \subset N_{\alpha} \otimes A
\label{cov.def.2}
\end{equation}
i.e. that $N_\alpha$ is both left- and right-invariant. It is enough
to show this for elements $n_{\alpha} (a) \in N_{\alpha}$.
We have
\begin{eqnarray*}
\Delta _{L}^{U} n_{\alpha} (a) &=& (\epsilon (\alpha) + \lambda) \sum
(a_{(1)} \otimes a_{(2)} \otimes 1 - a_{(1)} \otimes 1 \otimes
a_{(2)}) \\
& + &\sum ((S \alpha _{(2)}) \alpha _{(3)} a_{(1)} \otimes S
\alpha _{(1)} \otimes \alpha _{(4)} a_{(2)} - a_{(1)} ((S\alpha _{(2)})
\alpha _{(3)} \otimes a_{(2)} S \alpha _{(1)} \otimes \alpha _{(4)})) \\
& = & \sum a_{(1)} \otimes ((\epsilon (\alpha) + \lambda ) (a_{(2)}
\otimes 1 - 1 \otimes a_{(2)})\\
& + & S\alpha _{(1)} \otimes \alpha _{(2)}
a_{(2)} - a_{(2)} S \alpha _{(1)} \otimes \alpha _{(2)}) \\
& = & \sum a_{(1)} \otimes n_{\alpha} (a_{(2)}) \in A \otimes
N_{\alpha}.
\end{eqnarray*}

To prove the second part of (\ref{cov.def.2}) we will need the
following equality
\begin{equation}
S \alpha _{(2)} \otimes \alpha _{(3)} \otimes (S\alpha _{(1)})\alpha
_{(4)} = S\alpha _{(1)} \otimes \alpha _{(2)} \otimes 1
\label{cov.def.3}
\end{equation}
which holds for any $Ad_{R}$-invariant element $\alpha \in A$. Indeed,
\begin{eqnarray*}
(S \otimes id \otimes id) (\Delta \otimes id) Ad_{R} (\alpha) & = &
\sum (S \otimes id \otimes id) (\Delta \otimes id) (\alpha _{(2)}
\otimes (S\alpha _{(1)}) \alpha _{(3)}) \\
& = & \sum (S \otimes id \otimes id) (\alpha _{(2)} \otimes \alpha
_{(3)} \otimes (S\alpha _{(1)}) \alpha _{(4)}) \\
& = & S \alpha _{(2)} \otimes \alpha _{(3)} \otimes (S \alpha _{(1)})
\alpha _{(4)}.
\end{eqnarray*}
On the other hand, $\alpha$ is $Ad_{R}$-invariant, hence
\begin{eqnarray*}
(S \otimes id \otimes id) (\Delta \otimes id) Ad_{R} (\alpha ) = (S
\otimes id \otimes id) (\Delta \otimes id) (\alpha \otimes 1) = \sum
S\alpha _{(1)} \otimes \alpha _{(2)} \otimes 1.
\end{eqnarray*}
Now we have
\begin{eqnarray*}
\Delta _{R}^{U} n_{\alpha} (a) &=& \sum ( (\epsilon (\alpha) +
\lambda)(a_{(1)} \otimes 1 \otimes a_{(2)} - 1 \otimes a_{(1)} \otimes
a_{(2)}) \\
&+& S\alpha _{(2)} \otimes \alpha _{(3)} a _{(1)} \otimes (S\alpha
_{(1)}) \alpha _{(4)} a_{(2)} - a_{(1)} S \alpha _{(2)} \otimes \alpha
_{(3)} \otimes a_{(2)} (S\alpha _{(1)}) \alpha _{(4)}) \\
&=&\sum ((\epsilon (\alpha) + \lambda) (a_{(1)} \otimes 1 - 1 \otimes
a_{(1)} \otimes a_{(2)} \\
&+& S\alpha _{(1)} \otimes \alpha _{(2)} a_{(1)} \otimes a_{(2)} -
a_{(1)} S \alpha _{(1)} \otimes \alpha _{(2)} \otimes a_{(2)}) \\
&=& \sum n_{\alpha } (a_{(1)}) \otimes a_{(2)} \in N_{\alpha } \otimes
A.
\end{eqnarray*}
Hence we have proved that $(\Gamma _{\alpha}, d = \pi_\alpha
\circ D)$ is a first order bicovariant differential calculus as stated.

To prove (\ref{cov.def.1}) it is enough to observe that
\begin{equation}
n_{\alpha} (a) = \lambda Da - \omega _{U} (\alpha) a + a \omega _{U}
(\alpha)
\label{cov.def.4}
\end{equation}
Now applying $\pi_\alpha$ to the both sides of (\ref{cov.def.4}) we obtain
(\ref{cov.def.1}). \hspace{1cm} $\Box$ \vspace{20pt}

In this way we can assign a bicovariant differential calculus
$(\Gamma_\alpha ,d)$ on $A$ to any $Ad_R$-invariant $\alpha \in A$.
The differential calculus $(\Gamma_\alpha ,d)$ is universal in the
following sense. Every bicovariant sub-bimodule $N \subset A^2$ such
that $N_\alpha \subset N$ defines a bicovariant differential calculus
$\Gamma = A^2 /N$ such that $d = \pi D$ is given by (\ref{cov.def.1})
(with $\pi_\alpha$ replaced by $\pi$). In other words $N_\alpha$ is
the smallest sub-bimodule of $A^2$ leading to the bicovariant
differential calculus with a derivative $d$ given by
(\ref{cov.def.1}).

Applying the map $(id \otimes \epsilon \otimes id)\Delta_L^U$ to
$N_\alpha$ we see that the right ideal $\difid _\alpha \in \ker \epsilon$
corresponding to $N_\alpha$ is of the form
\[\difid _\alpha = \{ r_\alpha (a) b ; a,b \in A \} \]
where
\[r_\alpha(a) = (\lambda + \epsilon (\alpha) - \alpha) (\epsilon (a) -
a). \]
Using properties of the counit and the fact that $\alpha$ is an
$Ad_R$-invariant element of $A$ one can easily show that $\difid _\alpha$ is
$Ad_R$-invariant. Using that $(id \otimes \epsilon \otimes id)\Delta_L^U$
is an isomorphism of $A$-bimodules one can then obtain an alternative proof of
Theorem~\ref{cov.prop}. From this point of view
Theorem~\ref{cov.prop} can be thought of as a corollary of
Theorem~1.8 of \cite{woronowicz3}.

The universality of $(\Gamma_\alpha ,d)$ stated above takes the
following form in terms of $\difid _\alpha$: If $\difid  \subset \ker
\epsilon$ is an $Ad_R$-invariant ideal of $A$ such that $\difid _\alpha
\subset \difid $, then $\difid $ induces a bicovariant differential
calculus with $d$ given by (\ref{cov.def.1}).

The remainder of the section is devoted to developing several examples.

\begin{ex}
$4D_{\pm}$ calculi on $SL_q(2)$. \rm Let $A = SL_q(2)$, generated by
$2\times2$ matrix $\vect = ({t^i}_j)_{i,j=1^2}$. We use conventions such that
${t^1}_1 {t^1}_2 = q {t^1}_2 {t^1}_1$ etc. As is well-known the
q-deformed trace of \vect,
\[\tr_q \vect = {t^1}_{1} + q^{-2}{t^2}_2 \]
is an $Ad_R$-invariant element of $A$. Put $\alpha = \tr_q \vect$. The
right ideal $\difid _\alpha$ is generated by the four elements
\[{r^i}_j = (\lambda +1 +q^{-2} - \alpha)({\delta^i}_j - {t^i}_j). \]
If we set $\lambda = \lambda_\pm$ where
\[\lambda_+ = (1-q)(q^{-3}-1) \]
\[\lambda_- = -(1+q)(q^{-3} +1)\]
then we can extend $\difid _\alpha$ to the ideals $\difid _\pm$ which are
generated by ${r^i}_j$ and the five additional elements
\[({t^1}_2)^2 ,\quad {t^1}_2 ({t^2}_2 - {t^1}_1), \quad ({t^2}_1)^2,
\quad {t^2}_1({t^2}_2 - {t^1}_1), \]
\[q^2({t^2}_2)^2 - (1+q^2)({t^1}_1{t^2}_2 + q^{-1}{t^1}_2{t^2}_1) +
({t^1}_1)^2 .\]
The ideals $\difid _\pm$ generate the $4D_\pm$ calculi introduced in
\cite{woronowicz3}. Notice that the $4D_\pm$ calculi have a proper
classical limit when $q \rightarrow 1$ even though $\lambda_\pm
\rightarrow 0$ so that a singularity appears in the definition of $d$
in (\ref{cov.def.1}).  This is because the numerator also tends to zero in a
suitable way. This example suggests that (\ref{cov.def.1})
can be formulated more generally with $\lambda \in k$ provided that the
effects of the singular point can be cancelled in a suitable way.
\label{ex.4d} \end{ex}

\begin{ex} Bicovariant differential calculus on the quantum plane. \rm
Let $A = {\Bbb C}_q$, where ${\Bbb C}_q$ is the quantum plane generated
by $1$ and three elements $x, x^{-1},y$ modulo the relation $xy =
qyx$.  The bialgebra structure on $\Bbb C_q$ is given by $\Delta x^{\pm
1} = x^{\pm 1} \otimes x^{\pm 1}$, $\Delta y = y \otimes 1 + x \otimes
y$, etc. The element $x$ is $Ad_R$-invariant. Set $\alpha =x$. The
ideal $\difid _\alpha$ is generated by two elements:  \[(\lambda +1 -
x)(1-x), \quad (\lambda +1 - x)y.\] Now if we put $\lambda = q-1$ and
extend $\difid _\alpha$ to the ideal $\difid $ generated additionally
by $y^2$, then we obtain the bicovariant differential calculus
$III_{q,q^{-1}}$ described in $\cite{brzezinski4}$.  \label{ex.qplane}
\end{ex}

\begin{ex} Bicovariant differential calculi on $A(R)$. \rm This example
is a generalisation of Example~\ref{ex.4d}. Let $A(R)$ be the
bialgebra associated to a solution $R$ of the quantum Yang-Baxter
equation $R_{12}R_{13}R_{23} = R_{23}R_{13}R_{12}$, where $R_{12} =
R \otimes I$ etc. and $R \in M_n(k) \otimes M_n(k)$. The algebra of $A(R)$ is
generated by the matrix of generators $\vect =
({t^i}_j)_{ij=1}^{n} $ modulo the relation \cite{frt1} $R \vect_1
\vect_2 = \vect_2 \vect_1 R $. It has a standard coalgebra
structure induced by matrix multiplication. This $A(R)$ is usually made into a
Hopf algebra by a taking suitable determinant-like quotient. $R$ need
not be one of the standard $R$-matrices but we assume that it is
regular in the sense that such a quotient can be made to give an honest
(dual-quasitriangular) Hopf algebra $A$. In particular, we assume that
$R^{-1}$ and $\tilde R=((R^{t_2})^{-1})^{t_2}$ exist, where $t_2$
denotes transposition in the second matrix factor. Dual-quasitriangular
means that $R$ extends to a functional $\CR:A\tens A\to k$ obeying
axioms dual to those for a universal $R$-matrix. In particular, it obeys
\eqn{dqua}{ \CR({t^i}_i,{t^k}_l) = R^{i \; k}_{\;j \; l}, \quad \CR({t^i}_j,
S{t^k}_l) = {\tilde {R}^{i \; k}}_{\;j \;l}} while its extension to
products is as a bialgebra bicharacter (i.e.,
$\CR(ab,c)=\sum\CR(a,c\o)\CR(b,c\t)$ and
$\CR(a,bc)=\sum\CR(a\o,c)\CR(a\t,b)$).

In this context it has been shown in \cite{majid3} that $A(R)$ can be
transmuted to the bialgebra $B(R)$ living in the braided category of
$A(R)$-modules. The algebra of $B(R)$ is generated by the matrix $\vecu
$ with relations
\eqn{bgrel}{R_{21}\vecu_1 R_{12}\vecu_2=\vecu_2
R_{21}\vecu_1 R_{12}.}
These relations are known in various other
contexts also. The braided coalgebra structure is the standard matrix
one on the generators (extended with braid statistics).  Likewise for
their quotients: the Hopf algebra $A$ transmutes to a braided group $B$
living in the category of $A$-comodules. We need now the precise
relation between the product in the braided $B(R)$ and the original
$A(R)$ by which (\ref{bgrel}) were obtained (in an equivalent form) in
\cite{majid3}. This is given at the Hopf algebra level
by the transmutation formula
\eqn{trans}{a \underline \cdot b = \sum
a_{(2)} b_{(3)} \CR(a_{(1)},b_{(2)})\CR(a_{(3)}, Sb_{(1)})} where
$\underline\cdot$ denotes the product in $B$ while the right hand side
is in $A$. Here $B=A$ as a linear space,  as a coalgebra and as an
$A$-comodule under the adjoint coaction $Ad_R$\cite{majid3}.

We next define a matrix $\vartheta = ({\vartheta^i} _j)_{i,j=1}^n$,
where ${\vartheta ^i}_j = {\tilde {R} ^{i \; k}}_{\; k \; j}$
(summation over repeated indices). It has been shown in \cite{majid4}
that \begin{equation} \alpha_k = \tr (\vartheta \vecu^k), \quad
k=1,2,\ldots \label{alpha.k} \end{equation} are bosonic (i.e., invariant)
central elements of $B(R)$ and $B$. The product in (\ref{alpha.k}) is the
braided one in $B(R)$. For the standard $R$-matrices this recovers the
known Casimirs\cite{frt1} of the enveloping algebras $U_q(g)$, as
explained in \cite{majid4}. However, we are not limited to this case.

We now proceed to use (\ref{trans}) to view the $\alpha_k$ as elements
of $A(R)$ and $A$. The generators correspond, $\vecu=\vect$, while the
formula (\ref{trans}) is used to express the products of generators in
$B$ in terms of products in $A$ and evaluated using (\ref{dqua}) and
the bicharacter property for $\CR$, as explained in \cite[Sec.
5]{majid4}.  Computing $\vecu^n$ in this way, the first three
$\alpha_k$ immediately come out as
\align{ \alpha_1 &&= \tr (\vartheta
\vect)=\vartheta^i{}_jt^j{}_i\\
\alpha_2 &&= \tr (\vartheta R^{(1)}
\vect \vartheta R^{(2)} \vect)=\vartheta^i{}_j R^j{}_k{}^m{}_n
t^k{}_l\vartheta^l{}_m t^n{}_i\\
\alpha_3 &&= \tr (\vartheta R^{(1)}
R'^{(1)} \vect \tilde {R}^{(1)} \vartheta R^{(2)}R''^{(1)}\vect
\vartheta \tilde {R}^{(2)} R'^{(2)} R''^{(2)} \vect )\\
&&=\vartheta^i{}_j R^j{}_k{}^p{}_q R^k{}_l{}^w{}_y t^l{}_m {\tilde
R}^m{}_n{}^v{}_w \vartheta^n{}_p R^q{}_s{}^y{}_z t^s{}_u
\vartheta^u{}_v t^z{}_i}
where $R = R^{(1)} \otimes R^{(2)}$ and $R'$,
$R''$ are copies of $R$. One can compute all the $\alpha_k$ in a
similar way from (\ref{trans}).

Because these $\alpha_k$ are invariant elements of $B$, they are also
$Ad_R$-invariant elements of $A$. Hence we obtain a sequence of
bicovariant differential calculi on the quantum group $A$ obtained from
$A(R)$. The corresponding ideal $\difid _{\alpha_{k}}$ for fixed $k$ is
generated by $n^2$ elements \[ (\lambda_k + \epsilon(\alpha_k) -
\alpha_k)({\delta^i}_j - {t^i}_j), \quad \eps(\alpha_k)=\tr
\vartheta.\] Similarly, $\alpha$ given by any function of the
$\alpha_k$ will also do to define a bicovariant differential calculus,
which function can be chosen according to the needs of a specific
application.  Finally, for well-behaved $R$, these formulae can also be
used at the level of $A(R)$ itself, after this is made into a Hopf
algebra by formally inverting suitable elements.  \end{ex}
\vspace{12pt}

This completes the set of basic examples of differential calculi
obtained from $Ad_R$-invariant elements of $A$. We observe however that
if $\alpha , \beta \in A$ are $Ad_R$ invariant elements of $A$ then
$\alpha+\beta$ and $\alpha \beta$ are also $Ad_R$-invariant. So we have
a kind of `algebra' of differential calculi corresponding to the
algebra of $Ad_R$-invariant elements.

For example, if we find a single non-trivial $Ad_R$-invariant element
$\alpha$ of $A$ then we are able to define a hierarchy of
bicovariant differential calculi on $A$, i.e. $\Gamma_\alpha,
\Gamma_{\alpha^2}, \ldots$ We illustrate such a hierarchy by the
following:

\begin{ex}
Classification of bicovariant differential calculi on the line. \rm Let
us consider the Hopf algebra ${A} = k[x^{-1},x]$ with
comultiplication $\Delta x^{\pm 1} = x^{\pm 1} \otimes x^{\pm 1}$,
counit $\epsilon (x^{\pm 1}) =1$ and antipode $S(x^{\pm 1}) = x^{\mp
1}$. Clearly, all powers of $x$ and hence all the elements of $ A$
are $Ad_{R}$-invariant.  We put $\alpha=x$, $\lambda = q -1$ and
consider the differential calculi corresponding to
$\alpha,\alpha^2,\cdots$. For fixed $n>0$ the ideal $\difid _{\alpha
^n}$ is generated by the element \[(q-x^n)(1-x).\]
We notice that this ideal determines all the commutation relations in the
bimodule $\Gamma_{\alpha^n}$, which is now generated by $2n-1$ elements
$dx^{-n+1}, \ldots , dx^{-1},dx, dx^2, \ldots , dx^n$.  We have explicitly
\begin{eqnarray*}
xdx^{m} = dx^{m+1} - dxx^{m}, \hspace{1cm}  -n <m<n \\
xdx^{n} = dx^{n} x + (q^{-1}-1) dxx^{n}.\\
\end{eqnarray*}
\end{ex}
\begin{ex} \rm
Let $A = \Bbb C_q$ as in Example~\ref{ex.qplane}. We consider $\difid
_\alpha$ with $\alpha = x$ and we build the hierarchy $\difid
_{\alpha^n}$. Each $\difid _{\alpha^n}$ is generated by the two
elements \[(\lambda +1 - x^n)(1-x), \quad (\lambda+1 -x^n)y. \]
Assuming that $\lambda = q-1$ and adding one more generator $y^2$ we
can extend $\difid _{\alpha^n}$ to an ideal $\difid _n$. One can easily
see that the bimodule $\Gamma_n$ induced by $\difid _n$ is generated by
$2n$ elements $dy,dx^{-n+1},\ldots , dx^{-1},dx,dx^2, \ldots , dx^n$.
The commutation relations in $\Gamma_n$ read
\[xdx^n = dx^n x + (q^{-1} -1)dxx^n \]
\[xdx^m = dx^{m+1} -dx x^m , \quad -n<m<n \]
\[xdy = dyx \]
\[y dx^m = q^{-m}dx^m y + (q^{-1} -1)dy x^m, \quad -n<m \leq n \]
\[ydy = q^{-1}dyy. \]

\end{ex}

\end{document}